\documentclass[aps,prc,reprint,nofootinbib]{revtex4-1}
\usepackage{graphicx}   %       for graphics
\usepackage{latexsym}   %       for special symbols
\usepackage{enumerate}
\usepackage{tabularx}
\begin{document}
%==============================================================================
\title{First, second, third and fourth flow harmonics of deuterons and protons in Au+Au reactions at 1.23 A GeV}

\author{Paula Hillmann$^{1,2,3,4}$, Jan Steinheimer$^2$, Tom Reichert$^1$, Vincent Gaebel$^1$,  Marcus~Bleicher$^{1,2,3,4}$, Sukanya Sombun$^{5}$, Christoph Herold$^{5}$, Ayut Limphirat$^5$}

\affiliation{$^1$ Institut f\"ur Theoretische Physik, Goethe Universit\"at Frankfurt, Max-von-Laue-Str. 1, D-60438 Frankfurt am Main, Germany}
\affiliation{$^2$ Frankfurt Institute for Advanced Studies, Ruth-Moufang-Str. 1, 60438 Frankfurt am Main, Germany}
\affiliation{$^3$ GSI Helmholtzzentrum f\"ur Schwerionenforschung GmbH, Planckstr. 1, 64291 Darmstadt , Germany}
\affiliation{$^4$ John von Neumann-Institut f\"ur Computing, Forschungszentrum J\"ulich,
52425 J\"ulich, Germany}
\affiliation{$^5$School of Physics and Center of Excellence in High Energy Physics \& Astrophysics, Suranaree University of Technology, Nakhon Ratchasima 30000, Thailand}

\date{June 13, 2018}
\begin{abstract}
We explore the directed, elliptic, triangular and quadrangular flow of deuterons in Au+Au reactions at a beam energy of 1.23 AGeV within the UrQMD approach. These investigations are of direct relevance for the HADES experiment at GSI that has recently presented first data on the flow of light clusters in Au+Au collisions at 1.23 AGeV. To address the deuteron flow, UrQMD has been extended to include deuteron formation by coalescence. We find that this ansatz provides a very good description of the measured deuteron flow data, if a hard equation of state is used for the simulation. In addition we show that light cluster formation has a sizable impact on the proton flow and has to be taken into account to obtain reliable results in the forward/backward region. Based on the observed scaling of the flow, which is a natural result of coalescence, we conclude that deuteron production at GSI energies is a final state recombination effect. Finally, we also discuss the scaling relations of the higher order flow components up to $v_4$. We show that $v_3 \sim v_1v_2$ and $v_4 \sim v_2^2$ as function of transverse momentum and that the integrated $v_2^2 \sim v_4$ over the investigated energy range from $E_{lab}$=0.1 AGeV to 40 AGeV.
\end{abstract}

\maketitle

\section{Introduction}
Collisions of heavy ions in today's largest accelerators allow to explore the characteristics of nuclear matter under extreme temperatures and densities. While at the high temperature frontier, the goal is to explore the properties of the Quark-Gluon-Plasma, at the high density frontier the exploration of the nuclear equation of state (EoS) is in the center of interest. Especially for the understanding of compact stellar objects, e.g. neutron stars, a detailed knowledge of the EoS is of utmost importance. The  density range under investigation at corresponding temperatures is 3-4 times higher than  the nuclear ground state density and therefore of special interest as one expects a phase transition from nuclear/hadronic matter to deconfined matter. One may even expect more exotic forms of matter as quarkyonic \cite{McLerran:2008ux} or color superconducting matter \cite{Ruester:2006aj}.  Information on the EoS can be rather directly obtained from the study of the expansion of the fireball. In a simplified picture, the (explosive) expansion is driven by the initial pressure and therefore links the pressure to the finally observable transverse momentum spectra and its anisotropy in the observed hadrons \cite{Hofmann:1976dy,Stoecker:1979mj,Stoecker:1986ci,Ollitrault:1997vz}. During the last 20 years the study of flow has been refined and the transverse expansion is now studied in terms of a Fourier decomposition, see \cite{Alt:2003ab,Liu:2000am,Adamczyk:2014ipa,Andronic:2004cp,Adamczyk:2012ku,Pinkenburg:1999ya,Adamova:2002qx,BraunMunzinger:1998cg,Partlan:1994vs,Jena:2011ax,Puccio:2019oyd} for an overview of the experimental activities and see \cite{Poskanzer:1998yz,Bleicher:2000sx,Voloshin:2002wa,Petersen:2006vm,Snellings:2011sz,Retinskaya:2012ky,Yan:2013laa,Nara:2016phs,Yin:2017qhg,Ivanov:2014ioa,Hillmann:2018nmd,Nara:2017qcg,Isse:2005nk,LeFevre:2016vpp,Wang:2018hsw} for the corresponding theoretical investigations. These flow components, called $v_n$, are the expansion coefficients of the Fourier-series of the transverse momentum distribution \cite{Poskanzer:1998yz}:
\begin{equation}
E \frac{{\mathrm d}^3 N}{{\mathrm d}^3 {p}} = 
\frac{1}{2\pi} \frac {{\mathrm d}^2N}{p_{\mathrm{T}}{\mathrm d}p_{\mathrm T}{\mathrm d} y} 
  \left(1 + 2\sum_{n=1}^{\infty}  v_n \cos[n(\varphi-\Psi_{\rm RP})]\right), 
\label{fs}
\end{equation}
So one calculates the $v_n$ as average over all particles in a given event, accepting all events in the fixed centrality class \cite{Poskanzer:1998yz}:
\begin{equation}
v_n(p_{\mathrm{T}},y) = \langle \cos[n\varphi] \rangle.
\label{fc}
\end{equation}
Here, $\Psi_{\rm RP}$ denotes the reaction plane angle and $\varphi$ is the azimuthal angle with respect to the reaction plane. For the present analysis we always set $\Psi_{\rm RP}=0$ as given by the initial geometry of the system in the simulation. Until recently, higher flow components, e.g. the triangular flow, have only been studied as a consequence of initial state fluctuations at RHIC and LHC energies. 

The HADES experiment at the SIS18 accelerator at GSI in Darmstadt has measured Au+Au collisions at a fixed target beam energy of 1.23 AGeV, collecting a significant amount of data, sufficient to determine not only $v_1$ and $v_2$, but also higher flow components with the charged projectile spectators event plane method~\cite{Kardan:2016uog,Kardan:2017knj,Kardan:2017qau,Kardan:2018hna} which we approximate with the theoretical reaction plane. While initial state fluctuations (that drive the odd flow components at higher energies) are not connected to the participant reaction plane \cite{Adam:2016nfo,Esumi:2017qof,He:2017qsk,Krzewicki:2011ee,Alver:2010gr} one usually expected that triangular flow can not be observed with a fixed participant plane. However, in a recent study it was shown that this picture does not hold anymore for SIS beam energies \cite{Hillmann:2018nmd}. In particular at the beam energy under investigation in this study, a significant contribution to $v_3$, correlated with the reaction plane, was observed. At this energy one expects a substantial production of clusters allowing to explore their flow.

Just recently, new data on deuteron, triton and helium production at HADES energies (Au+Au reactions at the fixed target beam energy of 1.23 A GeV) has been published~\cite{Kardan:2018hna}. In this paper we present an analysis of this data for the directed, elliptic and triangular flow of deuterons. The focus of the current work is on the rapidity and transverse momentum spectra, and their scaling properties. For these studies, we use the UrQMD transport model with a hard EoS~\cite{Bass:1998ca,Bleicher:1999xi}.

\section{Simulation set-up}
The  Ultra relativistic Quantum Molecular Dynamics (UrQMD) transport model is applied to investigate the flow of deuterons. It is based on binary elastic and inelastic scattering of hadrons, resonance excitations and decays as well as string dynamics and strangeness exchange reactions~\cite{Bass:1998ca,Bleicher:1999xi,Graef:2014mra}. The model interprets scattering cross sections geometrically. If possible, the cross sections are taken from experimental data \cite{Patrignani:2016xqp}. For less known reactions effective model calculations, the additive quark model and detailed balance are used. For high beam energies $>8$ GeV, the mean particle production as well as collective flow in nuclear collisions can be well described by the cascade version of the transport model \cite{Bleicher:2005ti,Petersen:2006vm}. At lower beam energies as explored here nuclear and electromagnetic interactions can have a substantial effect on the dynamics and interactions of the particles. For the present investigation we use the same potentials as in our previous studies~\cite{Hillmann:2018nmd}.

We use a hadronic Skyrme-potential and the stiffness of the EoS (here a hard EoS) is given by $V_{Sk}$~\cite{Hartnack:1997ez}:\\
\begin{equation}
\centering
V_{Sk}=\alpha \cdot \left( \frac{\rho _{int}}{\rho _0}\right) +\beta \cdot \left( \frac{\rho _{int}}{\rho _0}\right) ^{\gamma}.
\end{equation}

\begin{table}[t]
\centering
\begin{tabular}{|c|c|}
\hline
Parameters & hard EoS \\
\hline
$\alpha$ [MeV] & -124\\
\hline
$\beta$  [MeV] & 71 \\
\hline
$\gamma$ & 2.00 \\
\hline
\end{tabular}
\caption{Parameters used in the UrQMD  Skyrme potential for a hard equation of state~\cite{Hartnack:1997ez}.\label{t1}}
\end{table}

By changing the parameters $\alpha$, $\beta$ and $\gamma$ one can change the stiffness of the nuclear equation of state. In the following we use the parametrization which gives us a rather 'hard' equation of state because this EoS led to a good description of the proton flows at the same energy~\cite{Hillmann:2018nmd}.
The parameters used for the hard equation of state in the present simulation are shown in table \ref{t1}.

We are aware that the inclusion of a momentum dependence in the nuclear potential~\cite{Li:2005gfa,Li:2005zza,Li:2007yd} may also allow to describe this data with a softer EoS. In addition one may include iso-spin dependent forces which have been discussed in the literature \cite{Li:2005gfa,Li:2007yd}. Here we stay with our previous set-up to explore whether proton flow and spectra and the collective flow of deuterons can be consistently described. 

For our investigation of the deuteron flows we supplement UrQMD with a coalescence approach as described in detail in \cite{Sombun:2018yqh}. Here one considers all possible proton-neutron pairs after their individual kinetic freeze-out (last scattering). The coalescence parameters in the two-particle restframe at equal times are: the relative coordinate distance $\Delta r=\left|r_p-r_n\right| <\Delta r_{max}=3.575$ fm$^{-3}$ and the relative momentum distance $\Delta r=\left|p_p-p_n\right| <\Delta p_{max}=0.285$ GeV.
This coalescence prescription has been shown to successfully describe deuteron production and spectra over a wide range of beam energies and system sizes, using only  this single set of parameters \cite{Sombun:2018yqh}.

\section{Results}
In this section results on the directed flow $v_1$, the elliptic flow $v_2$ and the triangular flow $v_3$ of deuterons and free protons in mid-peripheral gold-gold collisions at a fixed-target beam energy of 1.23 A GeV are presented. The value of the impact parameter in the simulations is fixed to $b$=6-9 fm, corresponding to a centrality of 20-30\%. 
The results shown are calculated employing a hard equation of state as discussed in the previous section. Spectator protons and neutrons are identified by having only soft collisions with hadrons inside their own nucleus and are removed from the analysis. Only deuterons made of two participating nucleons are considered for this study. The acceptance of the HADES detector has been implemented by using only hadrons with $18^{\circ}<\theta _{lab}<85^{\circ}$ for the analysis.

\subsection{The directed flow $v_1$}

\begin{figure}[t]	%       -----------------------------------------
\includegraphics[width=0.5\textwidth]{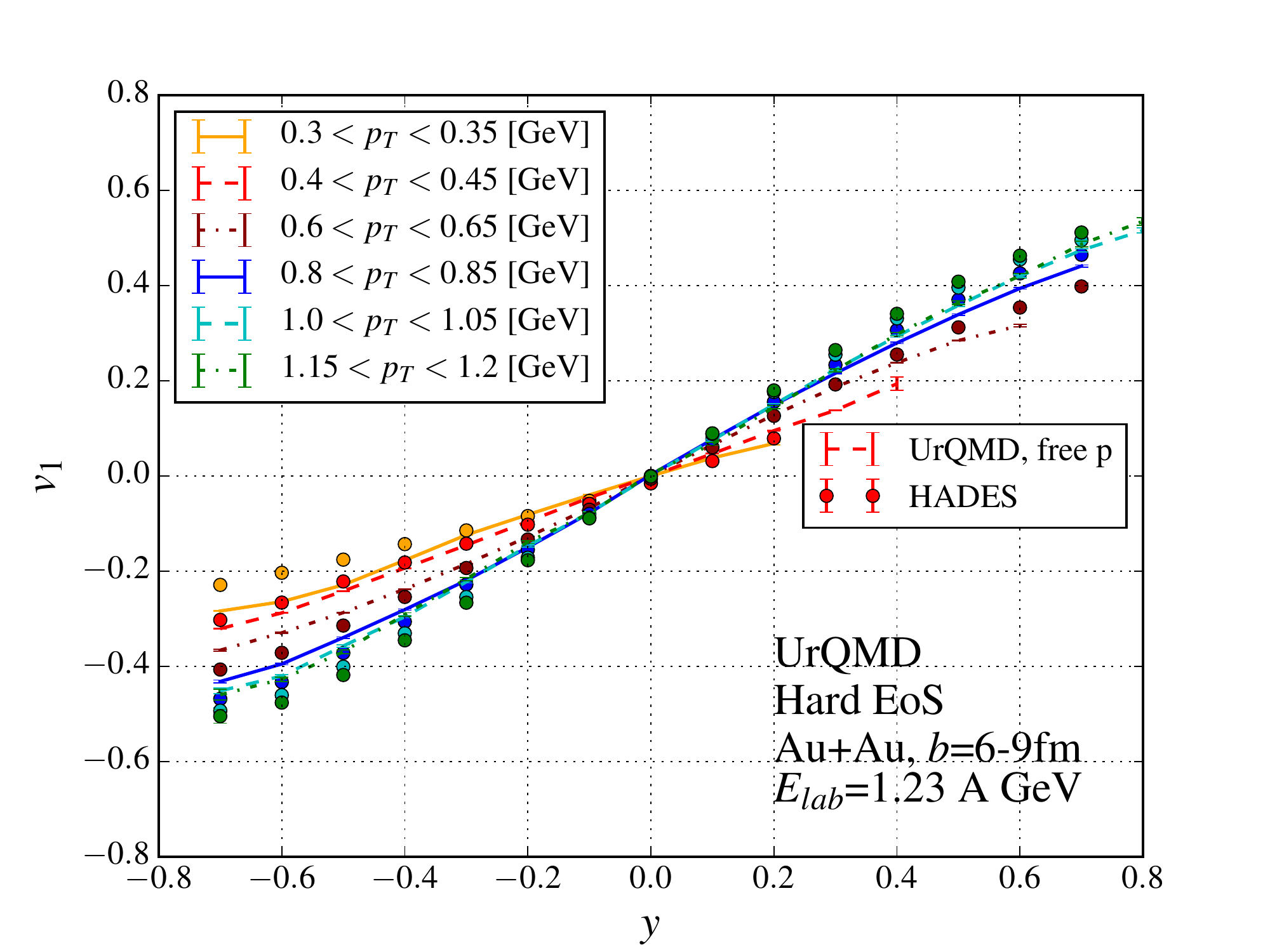}
\caption{[Color online] Directed flow of free protons in Au+Au collisions as a function of rapidity and for various transverse momentum regions at a fixed-target beam energy of 1.23 AGeV. The symbols denote the experimental data (20\%-30\% centrality)~\cite{Kardan:2017knj,Kardan:2017qau}, the lines indicate the UrQMD calculations ($b=6-9$ fm).}\label{f3}
\end{figure}	

Let us start by reexamining the directed flow of protons. In our previous studies, all protons (whether bound in clusters or not) were included in the flow analysis. The present calculation allows us to remove the protons in deuterons to obtain the 'free' protons to compare to the HADES data.

\begin{figure}[t]	%       -----------------------------------------
\includegraphics[width=0.5\textwidth]{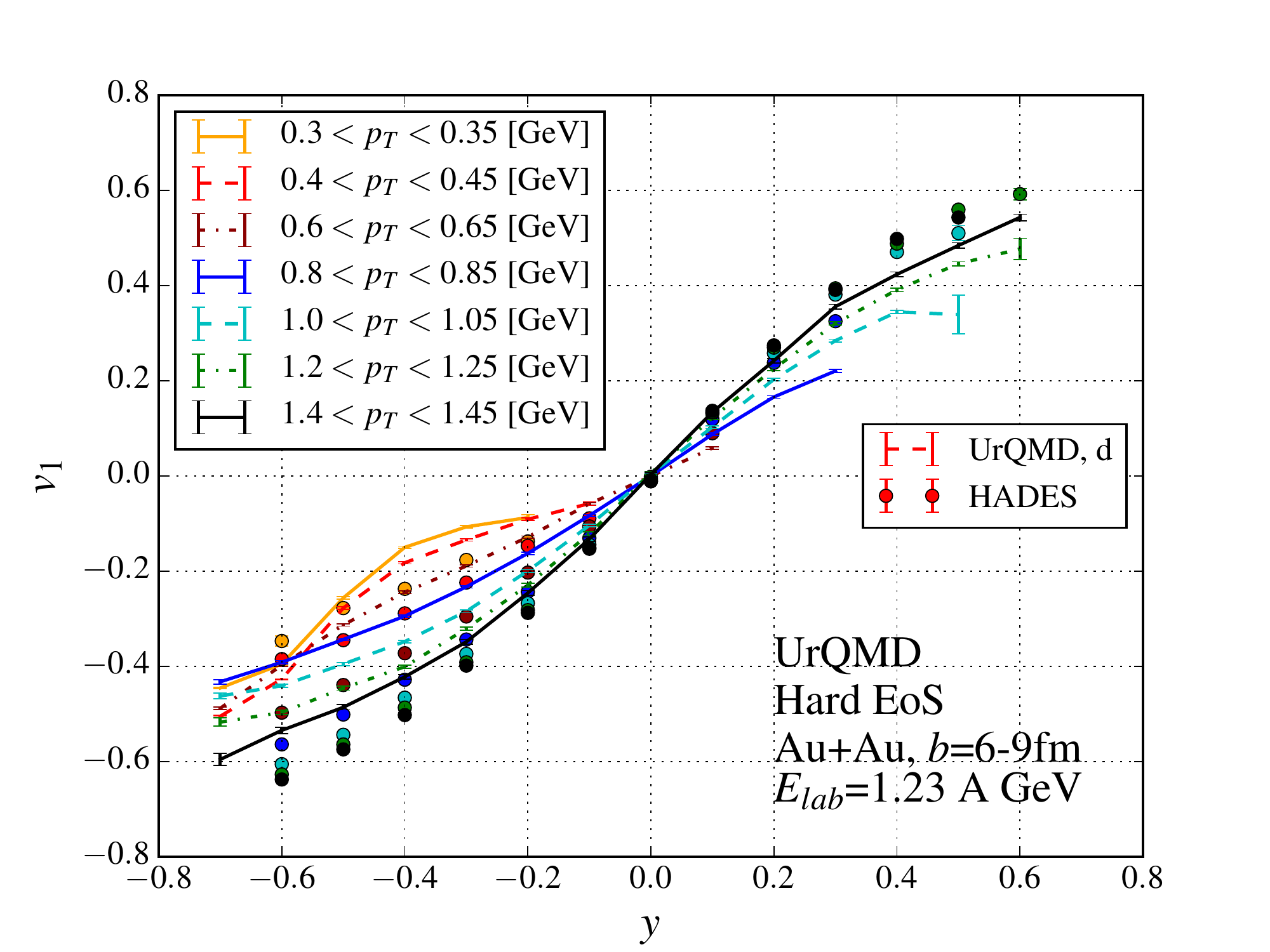}
\caption{[Color online] Directed flow of deuterons in Au+Au collisions as a function of rapidity and for various transverse momentum regions  at a fixed-target beam energy of 1.23 AGeV. The symbols denote the experimental data (20\%-30\% centrality)~\cite{Kardan:2018hna}, the lines indicate the UrQMD calculations ($b=6-9$ fm).}\label{f4}
\end{figure}

Figure \ref{f3} shows the directed flow of free protons as a function of rapidity for various transverse momentum regions. The lines denote the model calculations and the symbols show the preliminary HADES data~\cite{Kardan:2017knj,Kardan:2017qau}. In comparison to the results shown in \cite{Hillmann:2018nmd}, we observe an improved description of the low transverse momentum proton data if deuteron formation is taken into account. The directed flow of the deuterons is shown in Figure \ref{f4}.
As one can observe, the directed flow of deuterons follows that of the free protons  and shows a similar dependence on the transverse momentum. As expected, $v_1$ is most pronounced at large transverse momenta. As in the case of proton, the deuterons can be successfully described with a hard EoS. We have checked (not shown) that a good description of the data cannot be achieved by a soft momentum independent EoS. At very low transverse momenta the effect of contributing 'spectator' deuterons becomes visible. This deviation from the data at very low transverse momenta ($p_T < 0.45$ GeV) and near the fragmentation region is due to the unsatisfactory treatment of the nuclear break-up of the target/projectile remnants.

\begin{figure}[t]	%       -----------------------------------------
\includegraphics[width=0.5\textwidth]{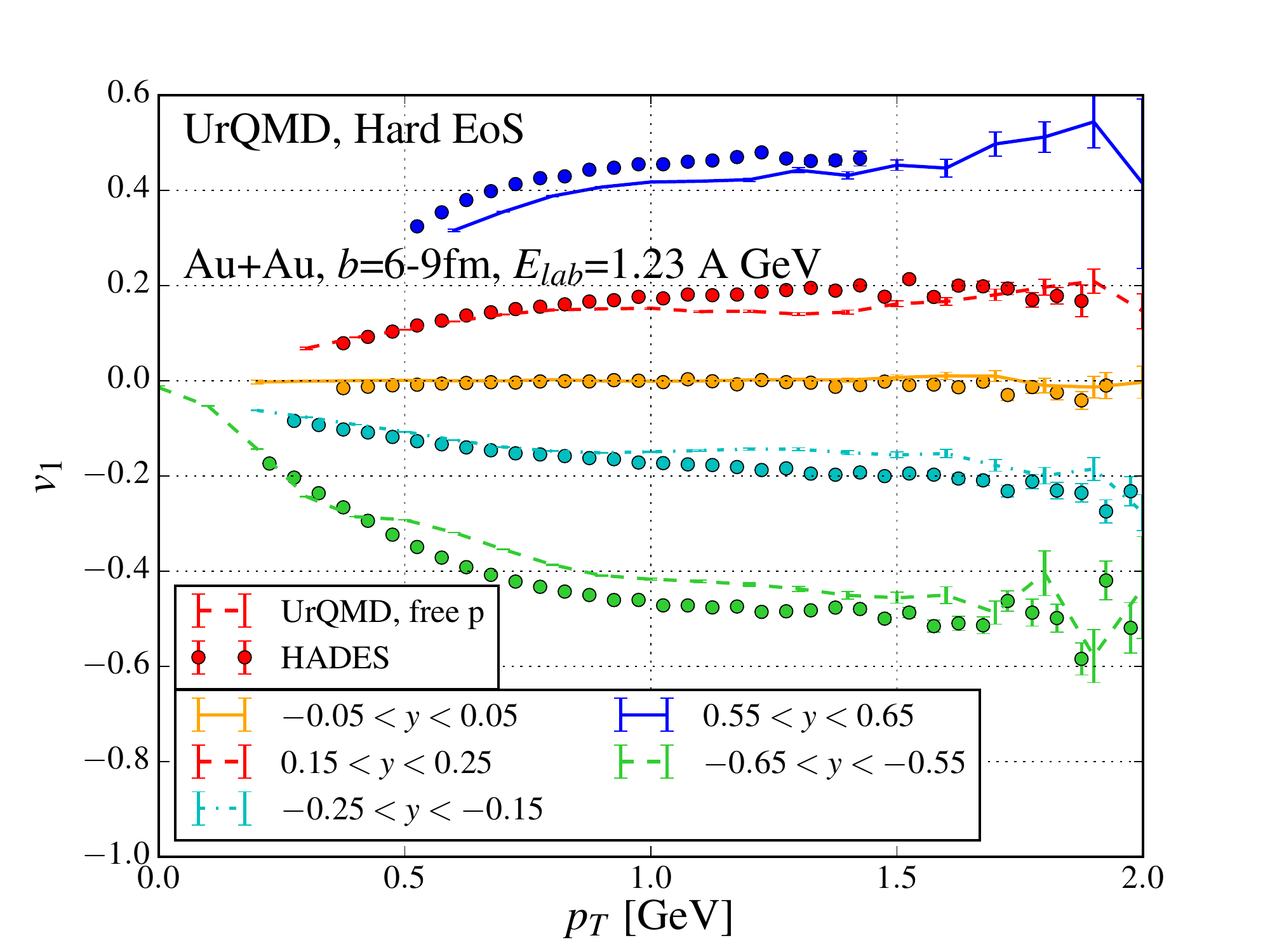}
\caption{[Color online] Directed flow of free protons in Au+Au collisions as a function of transverse momentum and for various rapidity regions at a fixed-target beam energy of 1.23 AGeV. The symbols denote the experimental data (20\%-30\% centrality)~\cite{Kardan:2017knj,Kardan:2017qau}, the lines indicate the UrQMD calculations ($b=6-9$ fm).}\label{f6}
\end{figure}	

\begin{figure}[t]	%       -----------------------------------------
\includegraphics[width=0.5\textwidth]{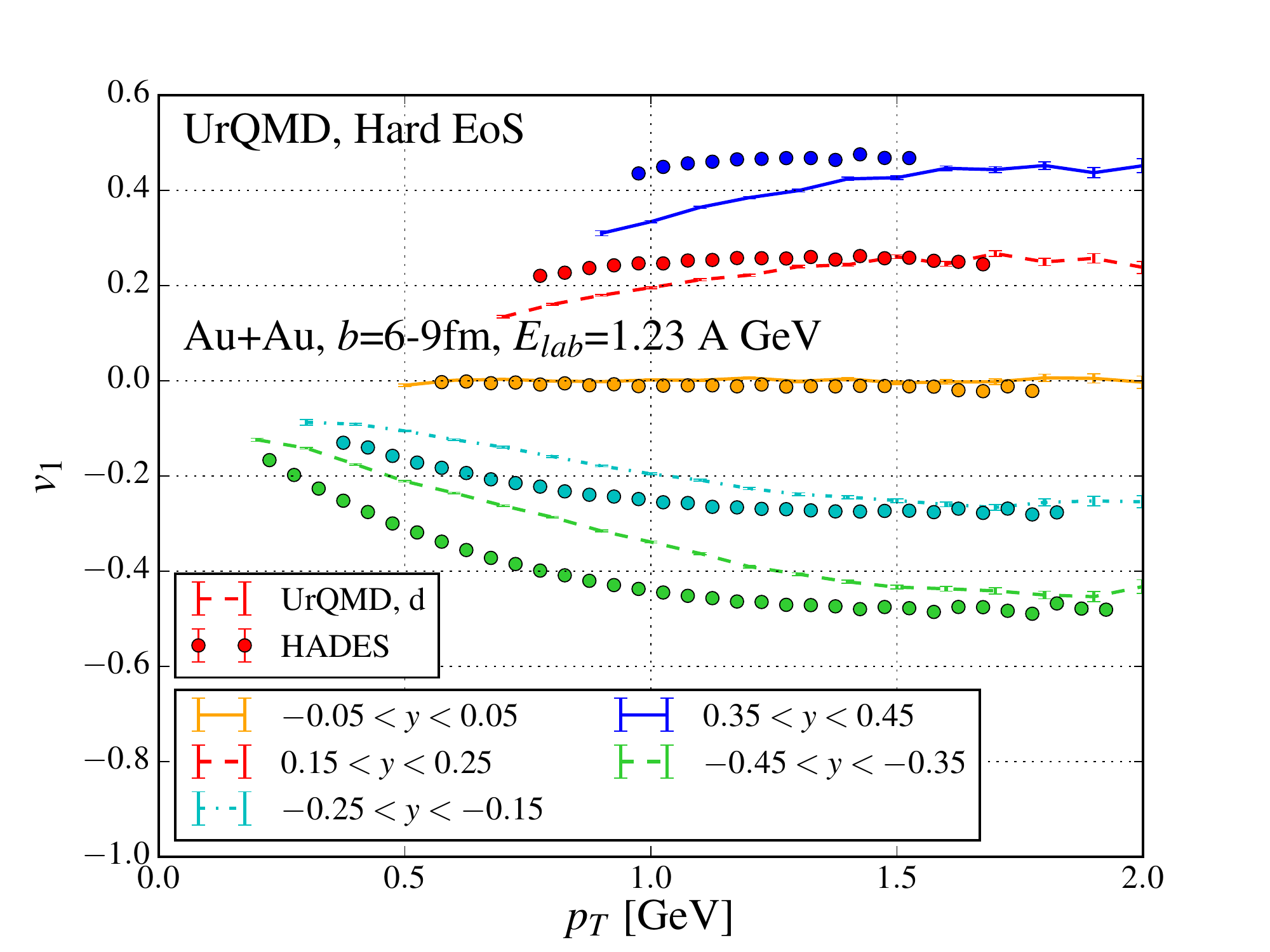}
\caption{[Color online] Directed flow of deuterons in Au+Au reactions as a function of transverse momentum and for various rapidity regions at a fixed-target beam energy of 1.23 AGeV. The lines indicate the UrQMD calculations ($b=6-9$ fm) and the data points show the preliminary HADES data (20\%-30\% centrality)~\cite{KardanTalk}.}\label{f7}
\end{figure}	

Next we explore the transverse momentum dependence in more detail. Figures~\ref{f6} and \ref{f7} show the directed flow $v_1$ as a function of transverse momentum for various rapidity bins for Au+Au collisions at a beam energy of 1.23 AGeV. The lines denote the UrQMD calculations ($b=6-9$ fm) and the symbols denote the preliminary experimental data (20\%-30\% centrality)~\cite{Kardan:2017knj,Kardan:2017qau,Kardan:2018hna}.
For the midrapidity window the value of the directed flow is exactly zero as expected due to momentum conservation for both deuterons and protons. In line with the rapidity dependence, towards forward rapidities, the dependence on the transverse momentum becomes stronger and the flow becomes more positive, towards the backward hemisphere the flow becomes negative. Generally, the proton flow is in good agreement with the experimental data. As expected from coalescence, the deuteron flow is more negative than the flow of the protons for all windows.
As already discussed for the rapidity dependence, the highest rapidity bin at low transverse momentum is less well described due to the incomplete treatment of the spectator region.

\subsection{The elliptic flow $v_2$}
The elliptic flow, corresponding to the second Fourier component $v_2$, is known as one of the most sensitive observables to the EoS. At center-of-mass energies above 10 GeV $v_2$ is typically the result of the free expanding almond-shaped overlap region in the direction of the y-axis (out-of-plane axis). In the low energy regime under investigation in this work $v_2$ is a consequence of the complex interplay of initial compression, a subsequent blocking of the in-plane emission by the spectators (leading to squeeze-out) followed by a final stage of in-plane expansion~\cite{Sorge:1998mk}. Depending on the EoS different time sections contribute with different strength and different sign to the final elliptic flow. At the energy discussed here, the major contribution stems from the intermediate phase. In this phase, the particles are mainly emitted out-of-plane which leads to negative $v_2$ with respect to the reaction plane.

\begin{figure}[t]	%       -----------------------------------------
\includegraphics[width=0.5\textwidth]{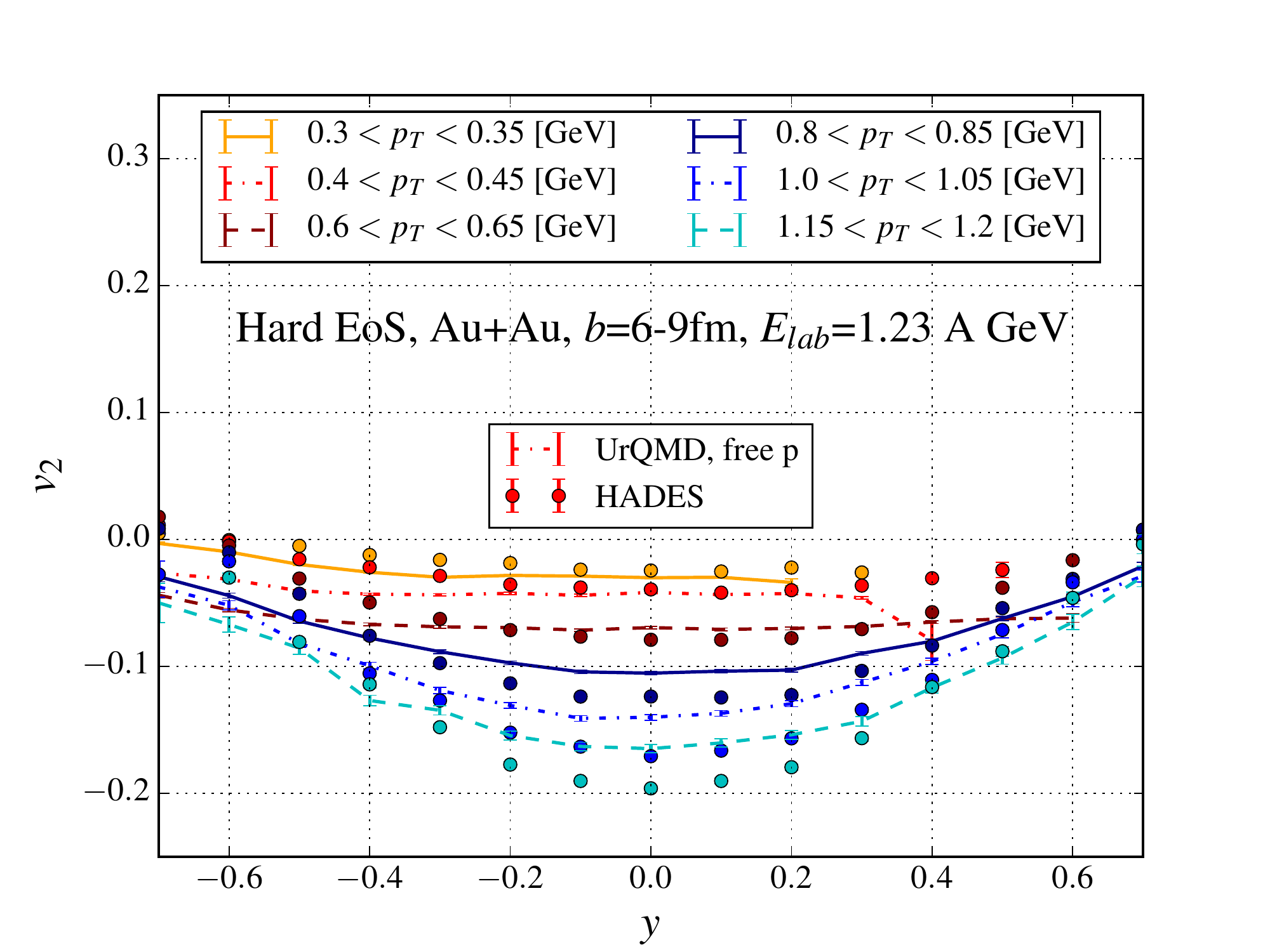}
\caption{[Color online] Elliptic flow of free protons in Au+Au collisions as a function of rapidity and for various transverse momentum regions at a fixed-target beam energy of 1.23 AGeV. The symbols denote the experimental data (20\%-30\% centrality)~\cite{Kardan:2017knj,Kardan:2017qau}, the lines indicate the UrQMD calculations ($b=6-9$ fm).}\label{f9}
\end{figure}	

\begin{figure}[t]	%     -----------------------------------------
\includegraphics[width=0.5\textwidth]{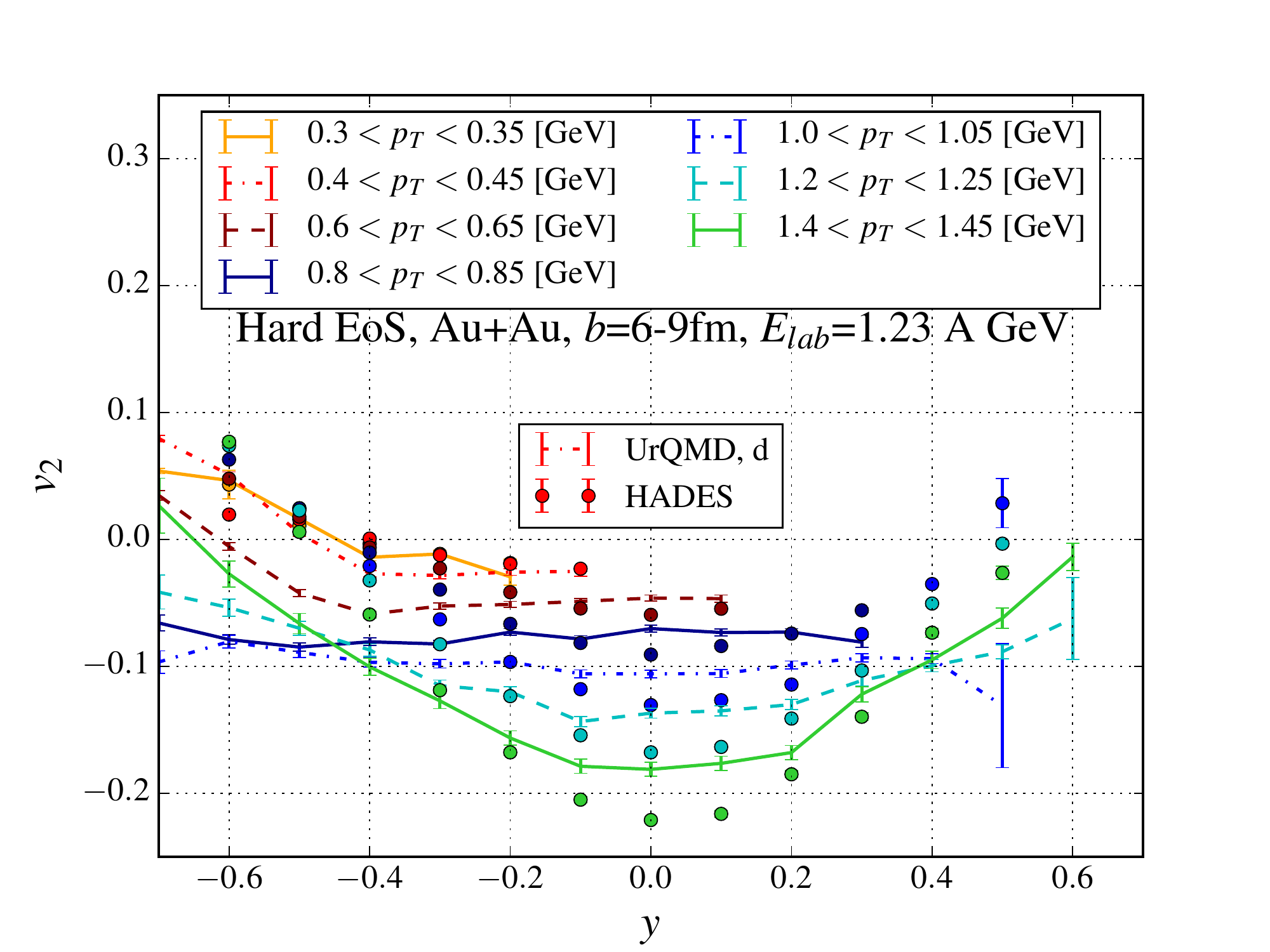}
\caption{[Color online] Elliptic flow of deuterons in Au+Au reactions as a function of rapidity and for various transverse momentum regions at a fixed-target beam energy of 1.23 AGeV. The lines indicate the UrQMD calculations ($b=6-9$ fm) and the symbols the preliminary HADES data (20\%-30\% centrality)~\cite{Kardan:2018hna}.}\label{f10}
\end{figure}	

Figures \ref{f9} and \ref{f10} show the elliptic flow $v_2$ of protons and deuterons as a function of rapidity for various transverse momentum regions at a fixed target beam energy of 1.23 AGeV. The lines denote the model calculations and the data points the preliminary HADES data~\cite{Kardan:2017knj,Kardan:2017qau,Kardan:2018hna}. 
For both protons and deuterons one observes a decrease of $v_2$ with increasing $p_T$ indicating a stronger absorption of high transverse momentum protons/deuterons, due to their earlier emission. The deuteron elliptic flow is larger than the elliptic flow of protons at any given $p_T$. This can be understood by the fact that deuterons are formed by coalescence which leads typically to $v_2^d\left( p_T^d\right) =2v_2^p\left( \frac{1}{2}p_T^d\right)$. We will explore this scaling further in Fig \ref{f13}.

\begin{figure}[t]	%       -----------------------------------------
\includegraphics[width=0.5\textwidth]{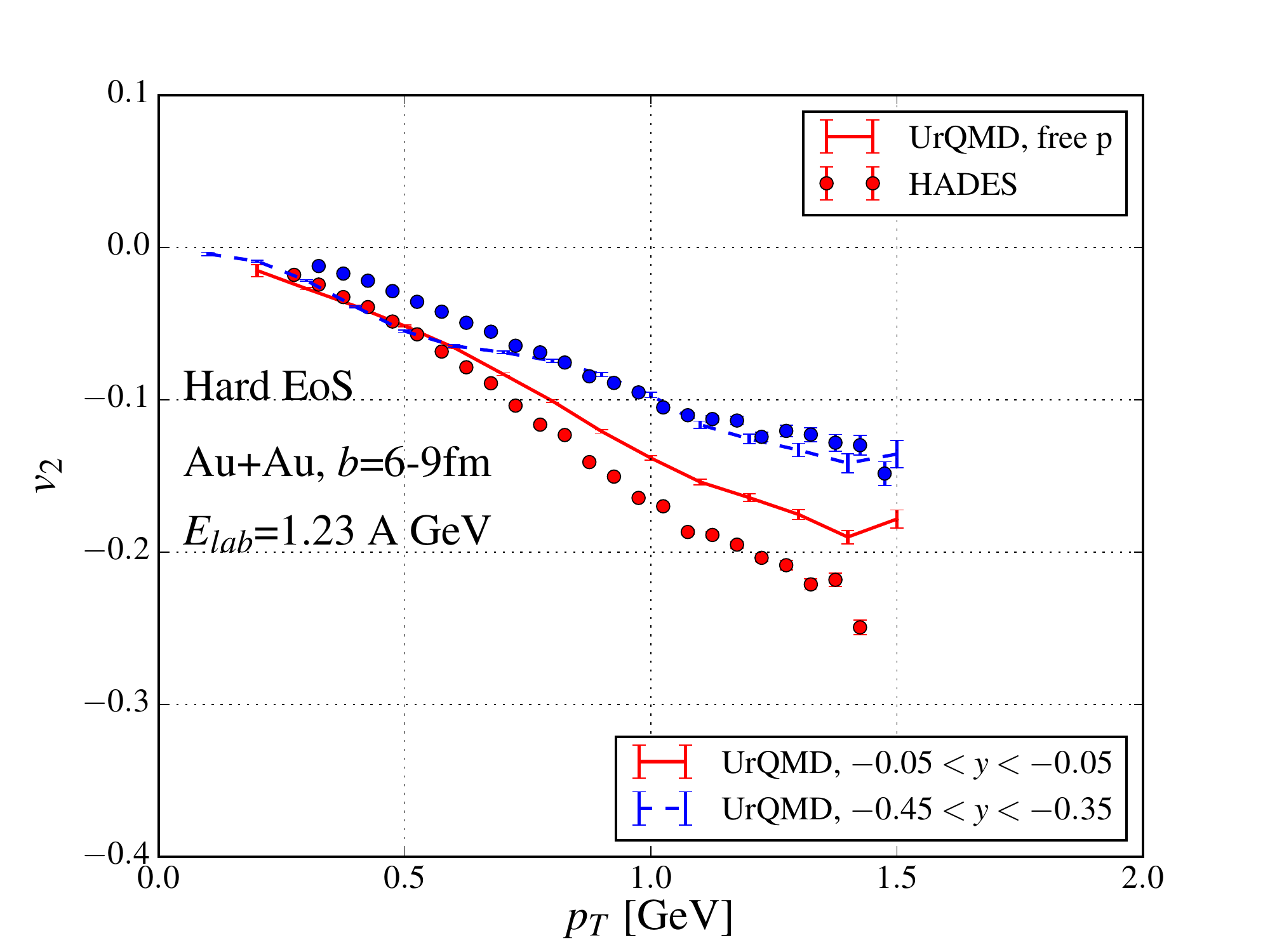}
\caption{[Color online] Elliptic flow of protons in Au+Au collisions as a function of transverse momentum and for various rapidity regions at a fixed-target beam energy of 1.23 AGeV. The symbols denote the experimental data (20\%-30\% centrality)~\cite{Kardan:2017knj,Kardan:2017qau}, the lines indicate the UrQMD calculations ($b=6-9$ fm).}\label{f11}
\end{figure}

\begin{figure}[t]	%       -----------------------------------------
\includegraphics[width=0.5\textwidth]{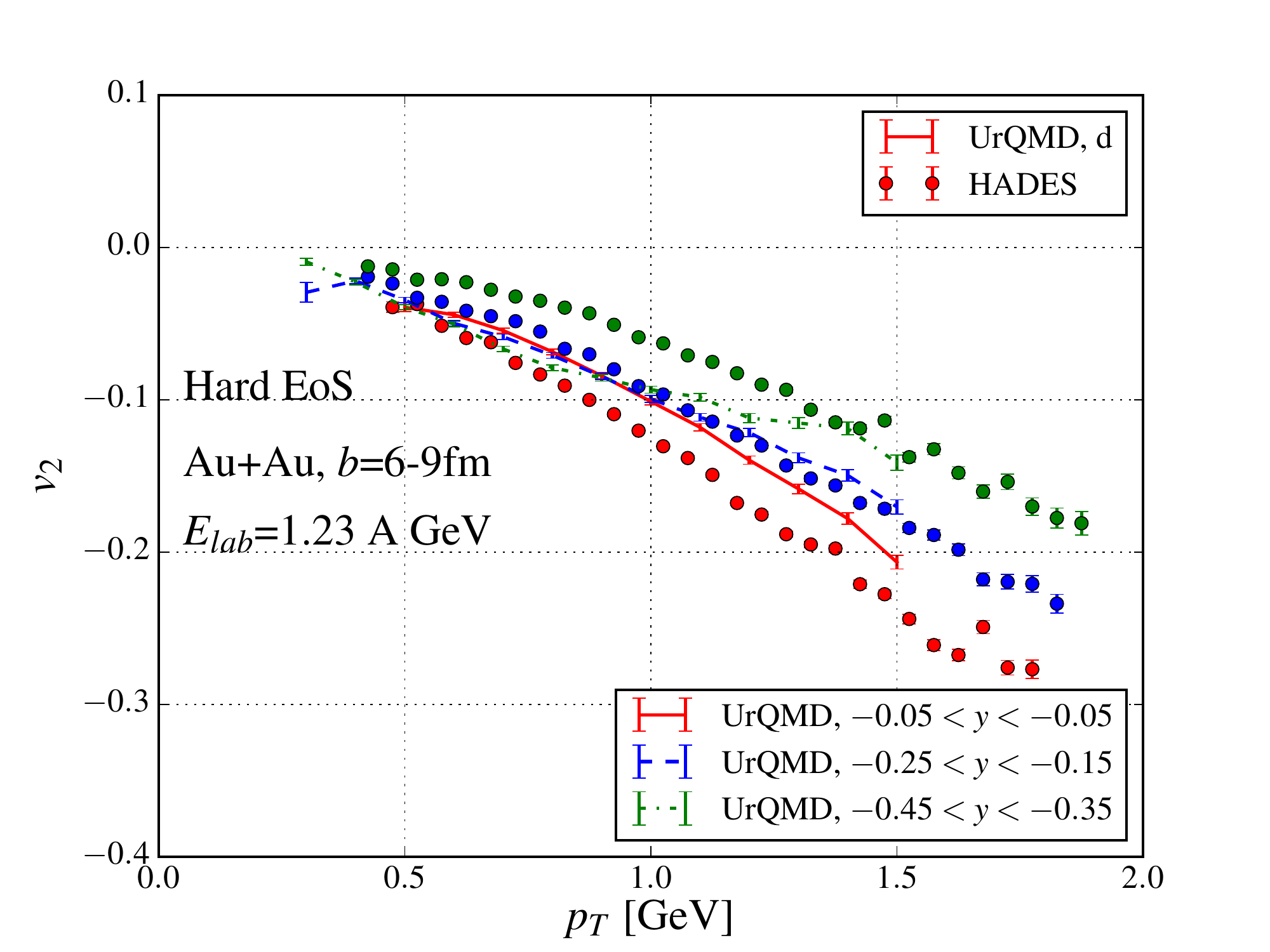}
\caption{[Color online] Elliptic flow of deuterons in Au+Au collisions as a function of transverse momentum and for various rapidity regions at a fixed-target beam energy of 1.23 AGeV. The lines indicate the UrQMD calculations ($b=6-9$ fm) and the symbols denote the preliminary HADES data (20\%-30\% centrality)~\cite{Kardan:2018hna}.}\label{f12}
\end{figure}

Next, we explore the transverse momentum dependence of the elliptic flow in Figs. \ref{f11} and \ref{f12}.

Figures \ref{f11} and \ref{f12} show the elliptic flow $v_2$ as a function of transverse momentum for different rapidity bins for Au+Au collisions at a beam energy of 1.23 AGeV. The lines denote the UrQMD calculations ($b=6-9$ fm) and the symbols denote the preliminary experimental data (20\%-30\% centrality)~\cite{Kardan:2017knj,Kardan:2017qau,Kardan:2018hna}.

One can observe that for both particles (protons and deuterons) the flow strongly increases with higher transverse momenta. For both investigated rapidity bins the calculations of the protons are in line with the experimental data. In the case of deuterons, we also observe a good agreement, except for the rapidity bin near the target region as discussed above.

\subsection{The triangular flow $v_3$}
Triangular flow is extremely interesting at low energies. The reason is that triangular flow is correlated to the event plane in the HADES energy region. This indicates an intricate interplay between different emission times. This is in strong contrast to high energies, where $v_3$ is only connected to initial state fluctuations and  not correlated to the reaction plane. This correlation of $v_3$ with the event-plane was first predicted in \cite{Hillmann:2018nmd} and confirmed by HADES data \cite{Kardan:2018hna}. It was pinned down to the time dependent interplay between the structure of the $v_2$ emission coupled with a strong $v_1$ component. 

\begin{figure}[t]	%       -----------------------------------------
\includegraphics[width=0.5\textwidth]{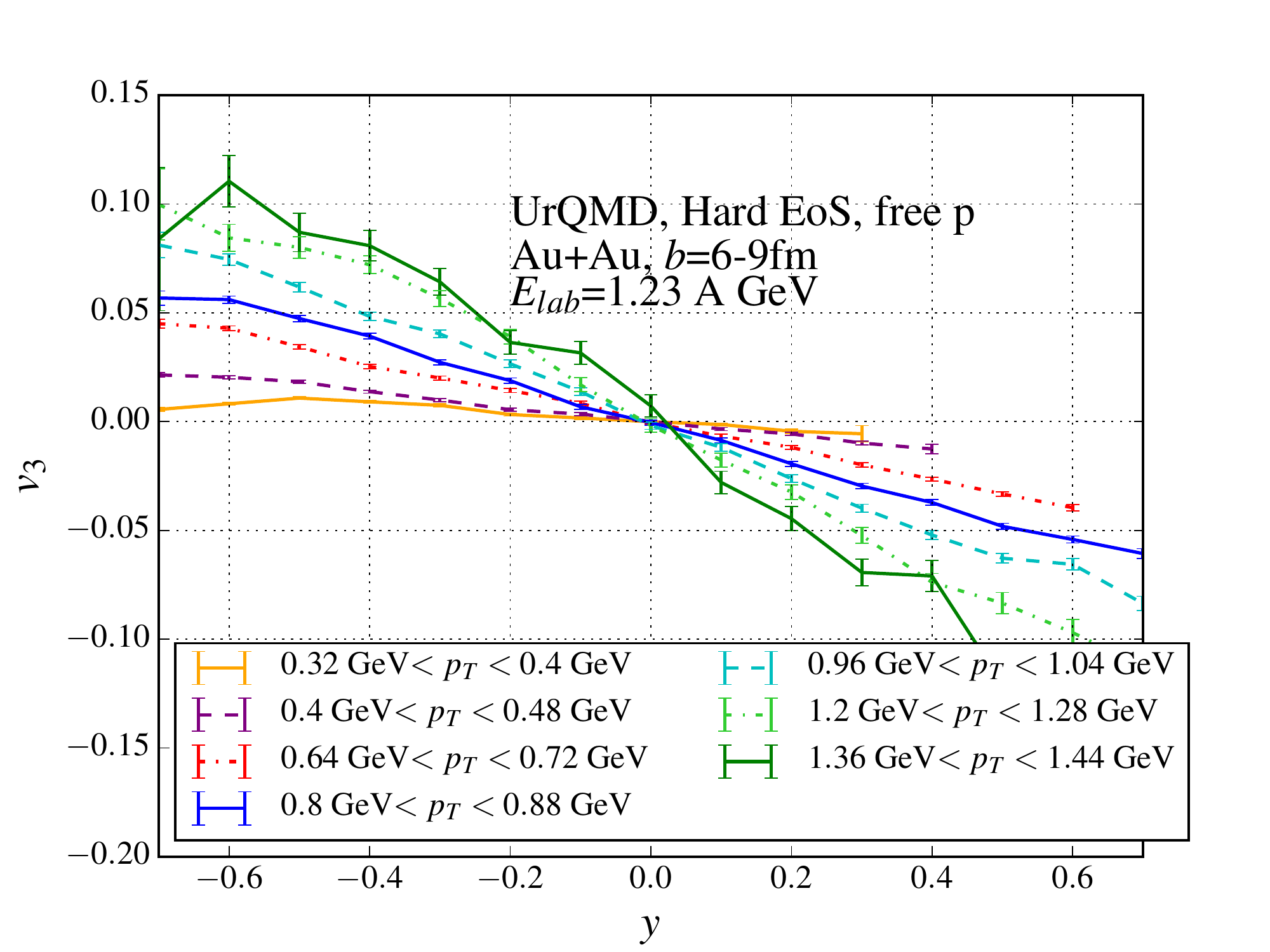}
\caption{[Color online] Triangular flow  of protons and deuterons in Au+Au collisions as a function of rapidity and for various transverse momentum regions at a fixed-target beam energy of 1.23 AGeV. The lines indicate the UrQMD calculations ($b=6-9$ fm).  
}\label{f14}
\end{figure}

\begin{figure}[t]	%     -----------------------------------------
\includegraphics[width=0.5\textwidth]{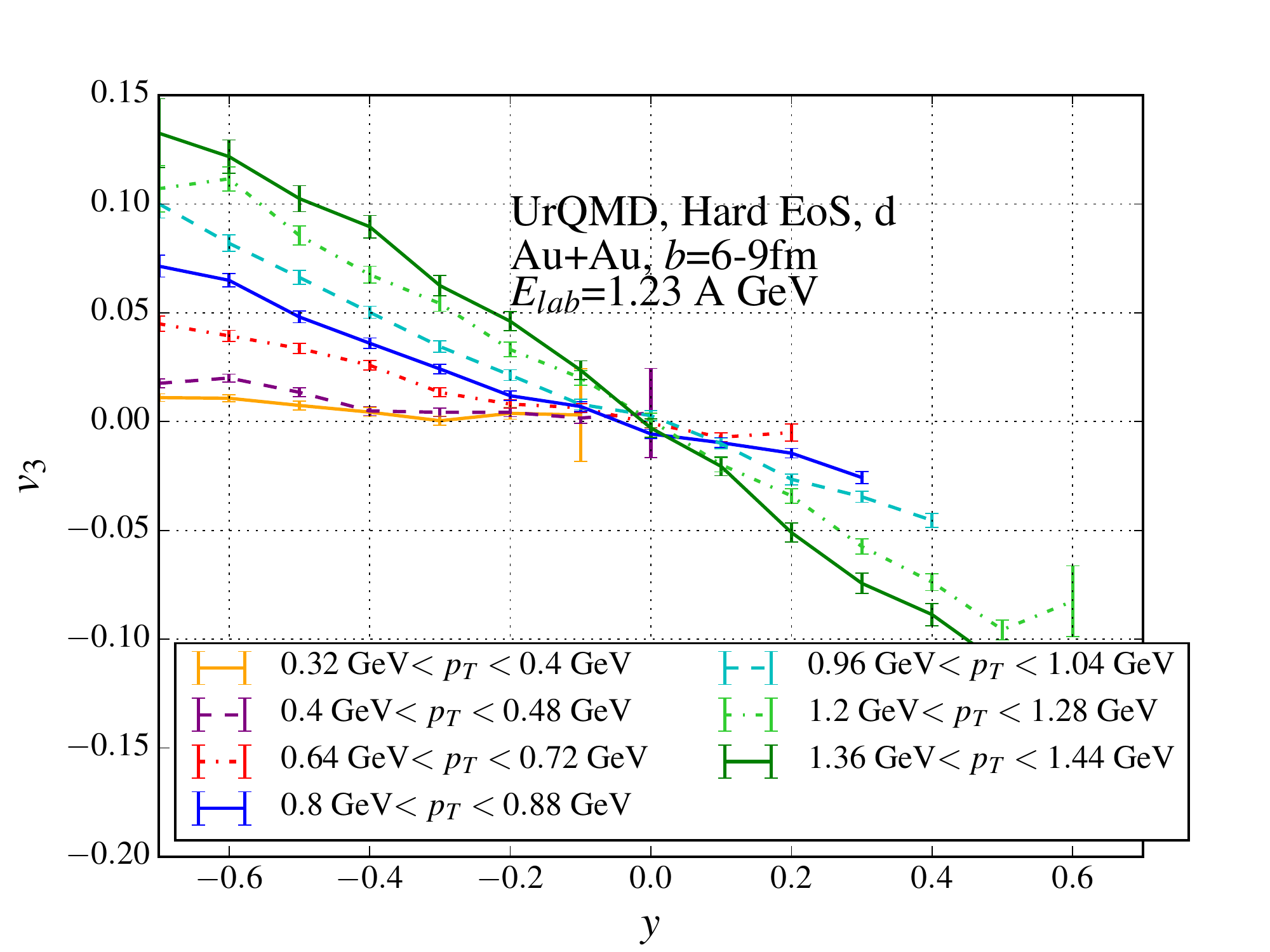}
\caption{[Color online] Triangular flow  of protons and deuterons in Au+Au collisions as a function of rapidity and for various transverse momentum regions. The lines indicate the UrQMD calculations ($b=6-9$ fm). 
}\label{f15}
\end{figure}

Figures \ref{f14} and \ref{f15} show the triangular flow of protons and deuterons as a function of rapidity for various transverse momentum regions in Au+Au collisions ($b=6-9$ fm) at a beam energy of 1.23 AGeV. The lines denote the UrQMD calculations.
Both protons and deuterons show an almost linear dependence on the rapidity for all transverse momentum windows. One can observe a $v_3\neq 0$ with respect to the reaction plane which is very similar for both particles and shows a strong rapidity dependence.

\begin{figure}[t]	%       -----------------------------------------
\includegraphics[width=0.5\textwidth]{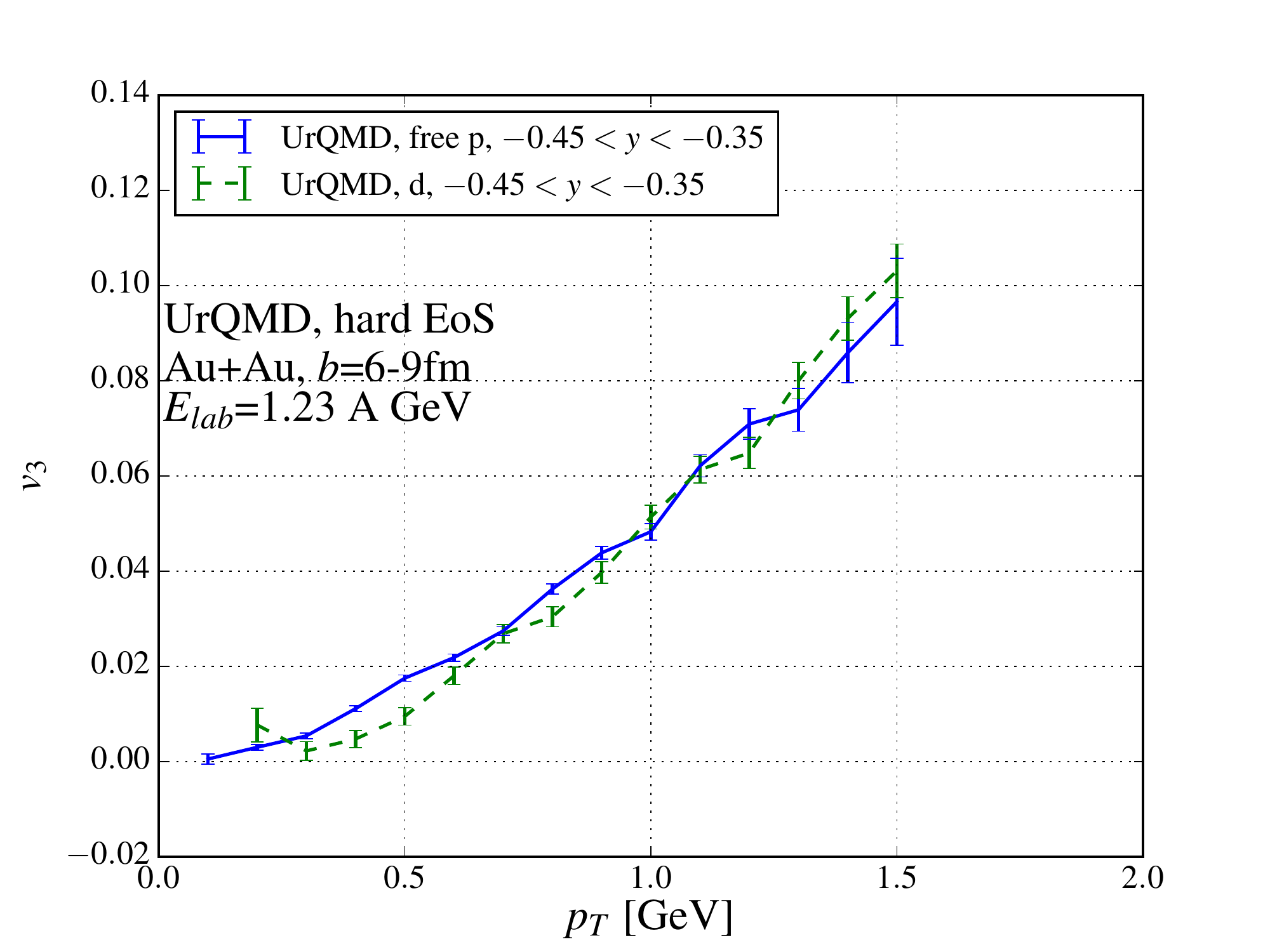}
\caption{[Color online] Triangular flow of free protons and deuterons in Au+Au collisions as a function of transverse momentum for the backward rapidity $-0.45<y<-0.35$ at a fixed-target beam energy of 1.23 AGeV. The lines indicate the UrQMD calculations ($b=6-9$ fm).}\label{f16}
\end{figure}	

\begin{figure}[t]	%       -----------------------------------------
\includegraphics[width=0.5\textwidth]{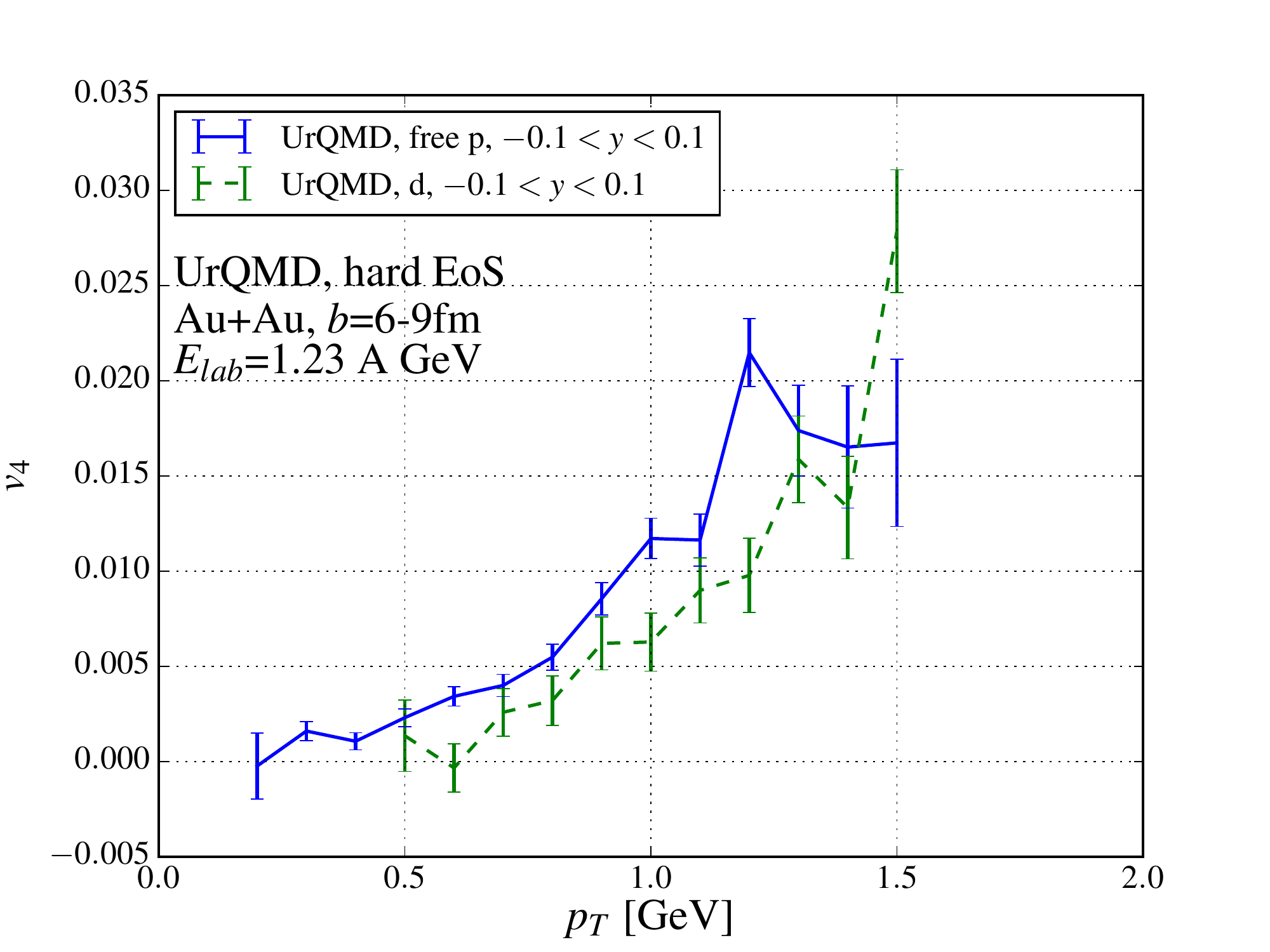}
\caption{[Color online] $4^{th}$ flow of free protons and deuterons in Au+Au collisions as a function of transverse momentum for the backward rapidity $|y|<0.1$ at a fixed-target beam energy of 1.23 AGeV. The lines indicate the UrQMD calculations ($b=6-9$ fm).}\label{f17}
\end{figure}	

Figure \ref{f16} shows the triangular flow of protons and deuterons as a function of transverse momentum for the backward-rapidity bin $-0.45<y<-0.35$ in Au+Au collisions ($b=6-9$ fm) at a beam energy of 1.23 AGeV. The lines denote the UrQMD calculations.
Both protons and deuterons show a strong increase of $v_3$ going to higher $p_T$. Surprisingly $v_3$ at HADES energy is on the same order as at RHIC energies~\cite{Adare:2011tg}. It is interesting to note that protons and deuterons show the same magnitude of $v_3$.

\subsection{$4^{th}$ order flow}
For the fist time a prediction of the $4^{th}$ order flow (quadrangular flow) with respect to the reaction plane is given for Au+Au reactions at 1.23 A GeV. Figure \ref{f17} shows the $4^{th}$ order flow of protons and deuterons as a function of transverse momentum for mid-rapidity $-0.1<y<0.1$ in Au+Au collisions ($b=6-9$ fm) at a beam energy of 1.23 A GeV. The lines denote the UrQMD calculations. Both protons and deuterons show a strong dependence on transverse momentum.

\section{Scaling and correlation analysis} 

\subsection{Mass number scaling}
\begin{figure}[t]	%       -----------------------------------------
\includegraphics[width=0.5\textwidth]{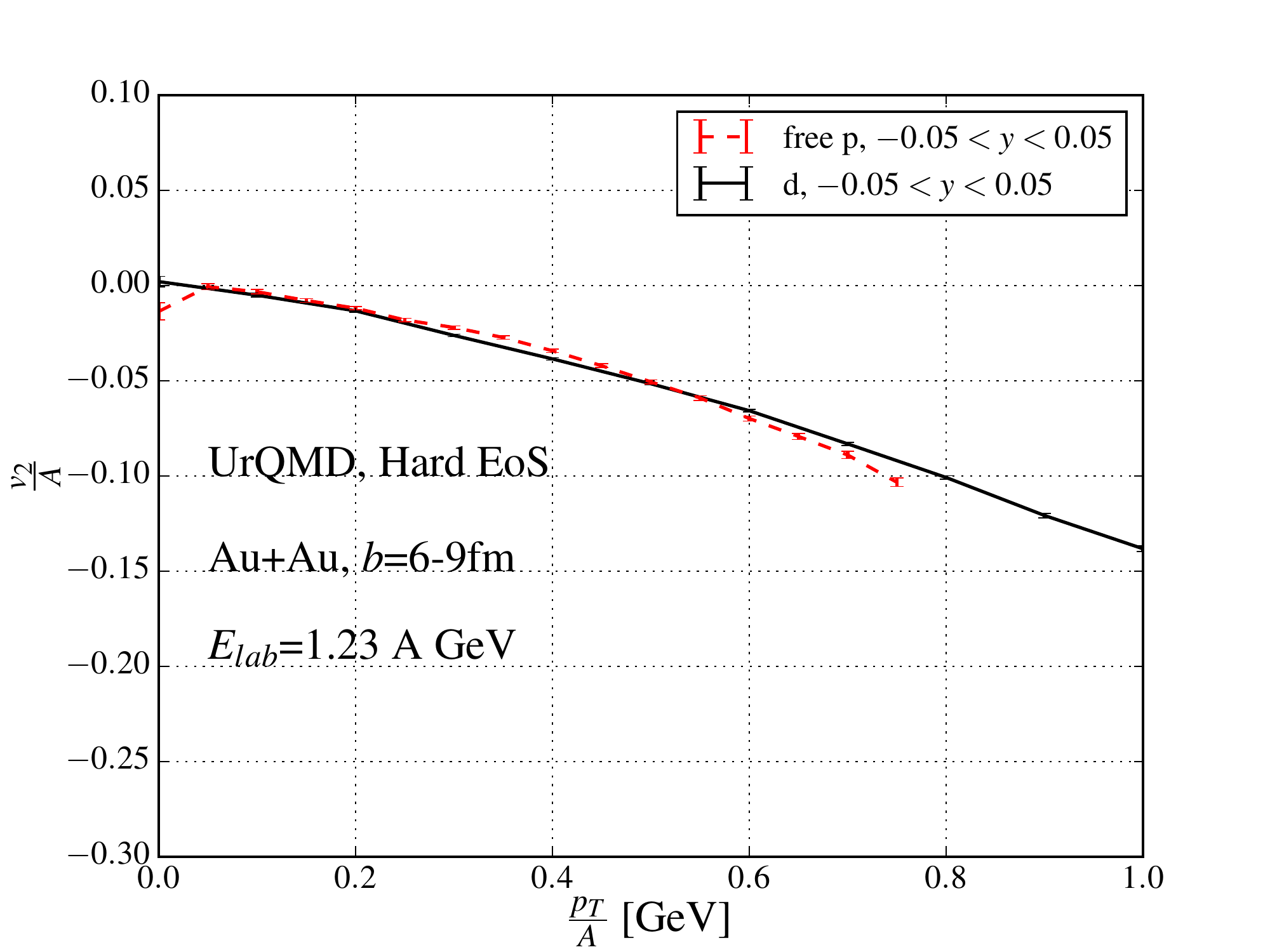}
\caption{[Color online] Elliptic flow of protons (solid line) and deuterons (dashed line) in Au+Au collisions as a function of transverse momentum and for $|y|<0.05$ scaled with the mass number $A$  at a fixed-target beam energy of 1.23 AGeV. The lines indicate the UrQMD calculations ($b=6-9$ fm). }\label{f13}
\end{figure}	

\begin{figure}[t]	%       -----------------------------------------
\includegraphics[width=0.5\textwidth]{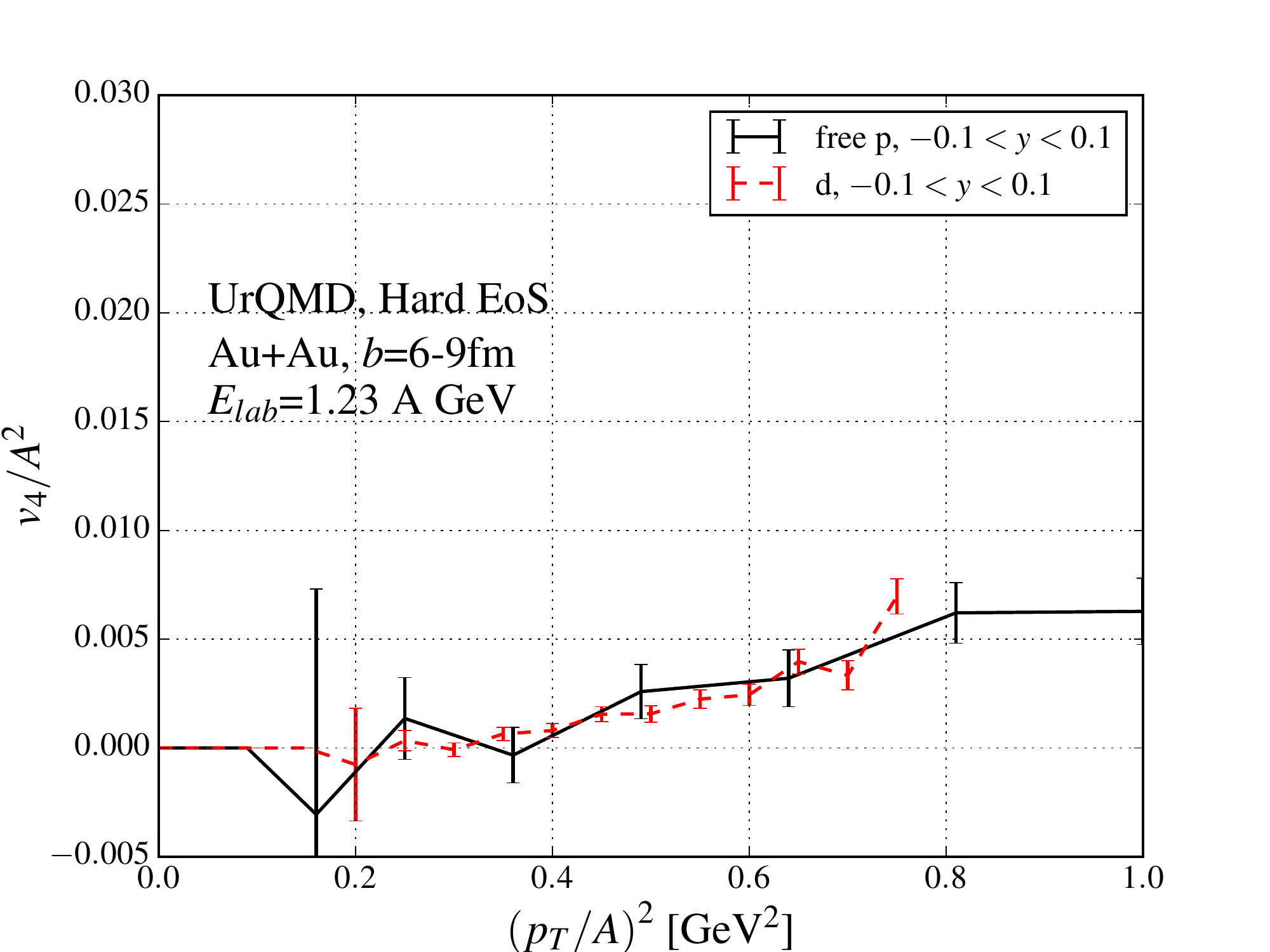}
\caption{[Color online] $v_4$ of protons (solid line) and deuterons (dashed line) in Au+Au collisions as a function of transverse momentum and for $|y|<0.1$ scaled with the mass number $A$  at a fixed-target beam energy of 1.23 AGeV. The lines indicate the UrQMD calculations ($b=6-9$ fm). }\label{f18}
\end{figure}	

The scaling of elliptic flow with the number of constituents has long been established with quark recombination models at RHIC energies~\cite{Fries:2003kq}. For the coalescence of nucleons into deuterons the same scaling is present in terms of the baryon number. This results in the expectation that $v_2^d\left( p_T^d\right) =2v_2^p\left( \frac{1}{2}p_T^d\right)$. Thus $v_2/A$ as function of $\left( p_T /A\right)$, with $A$ being the baryon number, should yield the same curves for protons and deuterons, if deuterons are formed by coalescence. Taking the data of Fig.~\ref{f11} and \ref{f12} we show the scaled flow of protons and deuterons for Au+Au collisions (20\%-30\% centrality) at a beam energy of 1.23 AGeV in Fig. \ref{f13}. We observe that the simulation predicts perfect scaling, as expected from the implemented coalescence mechanism. A confirmation of this mass number scaling would strongly support the idea of deuteron formation by coalescence and would disfavour direct emission of deuterons from the (thermal) fireball.

In Fig.~\ref{f18} we explore if such a scaling also translates to the quadrangular flow $v_4$. Here we scale with $A^2$. Also in this case scaling between deuterons and protons is present, indicating the tight connection between both particles. 

\subsection{Flow correlations} 
While constituent-number-scaling as discussed above provides insight into the formation of composite objects, the correlations between higher order flow coefficients can yield information on the underlying dynamics. A prime example in this respect has been the predicted scaling of $\frac{v_4\left( p_T\right) }{v_2^2\left( p_T\right)}$ as suggested in~\cite{Borghini:2005kd}. It was argued that for high transverse momenta $\left( T<< p_T\right)$ a universal result of $\frac{v_4}{v_2^2}=\frac{1}{2}$ emerges in the case of an ideal fluid expansion. At RHIC energies however, the experimentally observed ratio was $\frac{v_4}{v_2^2}\sim 1.4$~\cite{Adams:2003zg}.

\begin{figure}[t]	%       -----------------------------------------
\includegraphics[width=0.5\textwidth]{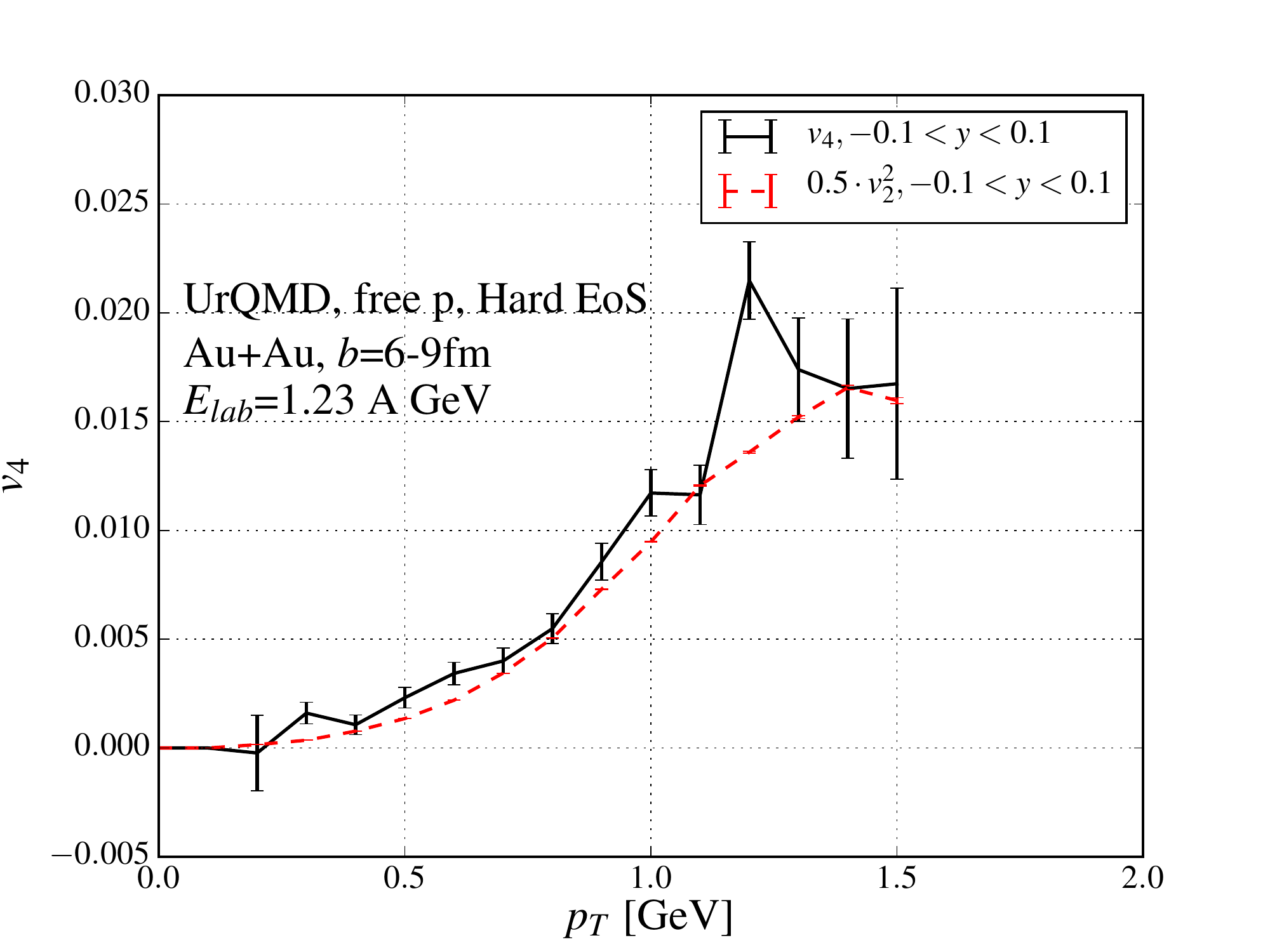}
\caption{[Color online] Flow of free protons in Au+Au collisions ($b=6-9$ fm) as a function of transverse momentum for $|y|<0.1$  at a fixed-target beam energy of 1.23 AGeV. The lines indicate the UrQMD calculations for a hard equation of state.}\label{f19}
\end{figure}

\begin{figure}[t]	%       -----------------------------------------
\includegraphics[width=0.5\textwidth]{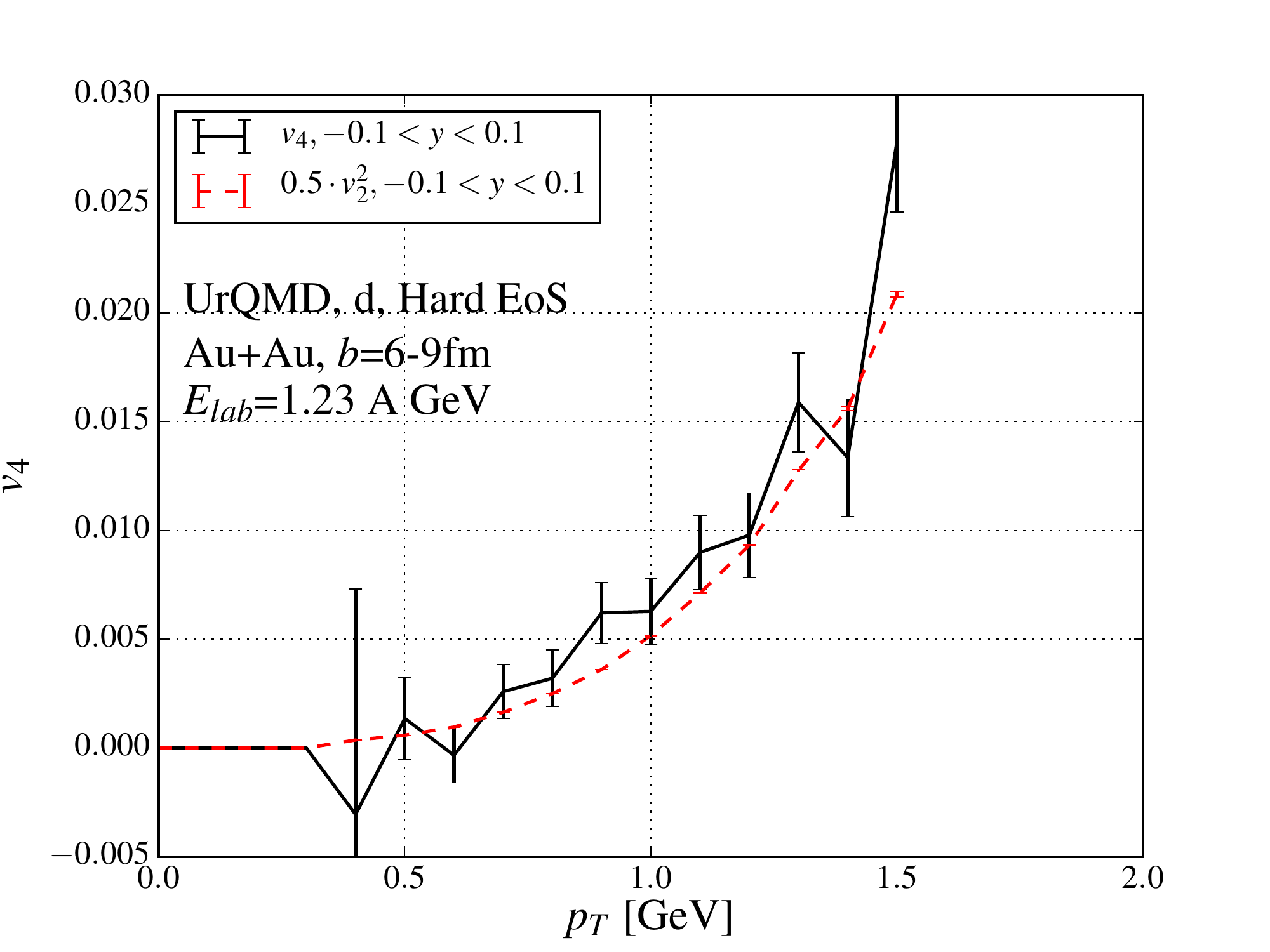}
\caption{[Color online] Flow of deuterons in Au+Au collisions ($b=6-9$ fm) as a function of transverse momentum for $|y|<0.1$  at a fixed-target beam energy of 1.23 AGeV. The lines indicate the UrQMD calculations for a hard equation of state.}\label{f20}
\end{figure}

In Figs. \ref{f19}-\ref{f20} we explore the scaling of $\frac{v_4}{v_2^2}$ in Au+Au reactions at $E_{lab}$=1.23 A GeV. In Fig. \ref{f19} we present a comparison between the free proton $v_4\left( p_T\right)$ and $\frac{1}{2}v_2^2\left( p_T\right)$ as function of transverse momentum. For the whole explored transverse momentum region up to $p_T$=1.5 GeV, we observe excellent scaling. In Fig. \ref{f20} we explore the same scaling for deuterons and also observe consistent scaling of $\frac{v_4}{v_2^2}=\frac{1}{2}$. In the light of the discussion above, the obtained results are in line with expectations from the expansion of an ideal fluid.

\begin{figure}[t]	%       -----------------------------------------
\includegraphics[width=0.5\textwidth]{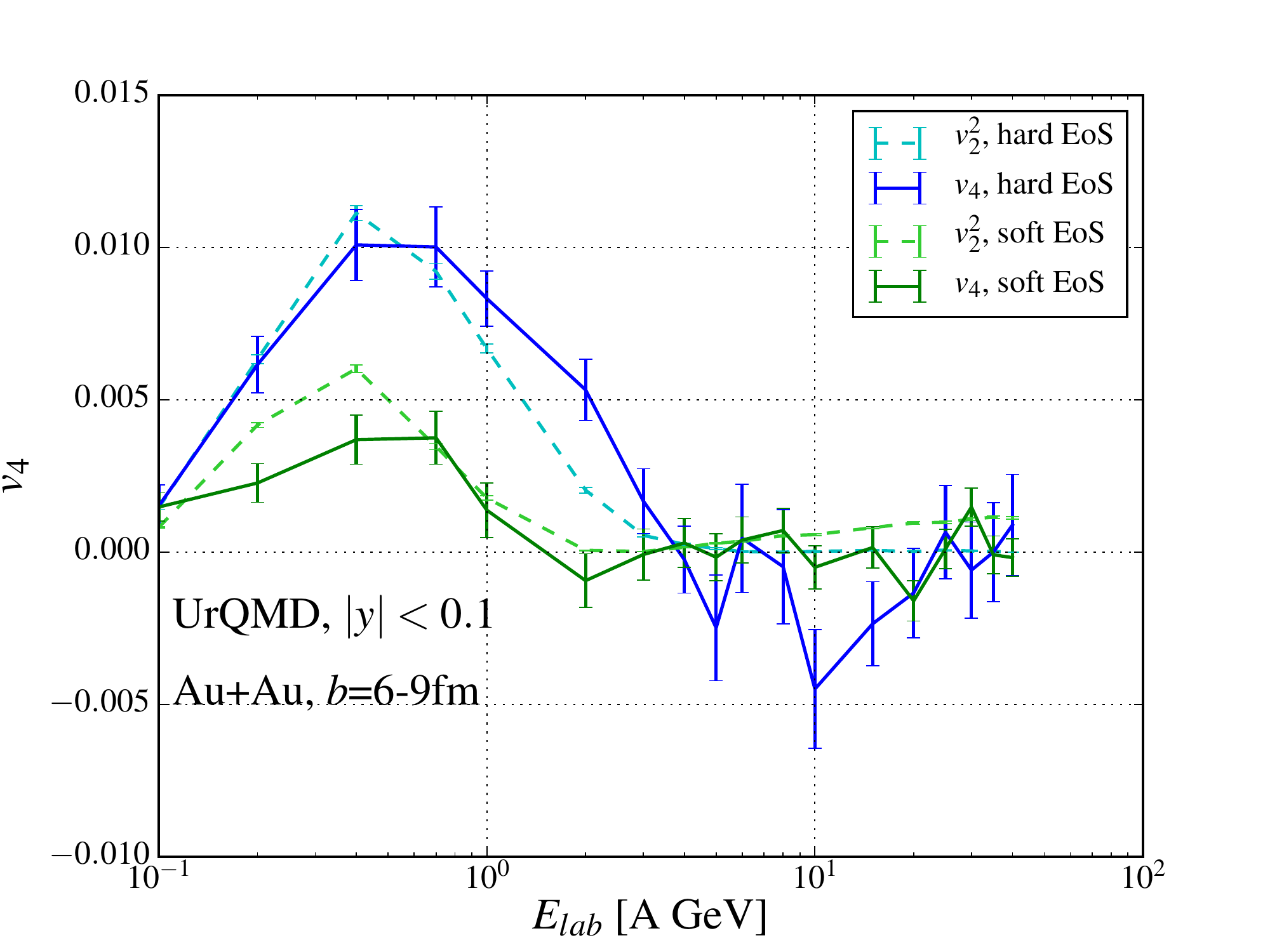}
\caption{[Color online] Flow of protons (solid line) in Au+Au collisions ($b=6-9$ fm) as a function of the beam-energy $E_{lab}$. The lines indicate the UrQMD calculations for a hard and soft equation of state. }\label{f21}
\end{figure}

A direct scaling is even present for the integrated $v_4$ and $v_2^2$ values at midrapidity. This is demonstrated in Fig. \ref{f21}, where we show $v_4$ and $v_2^2$ of protons at midrapidity as function of beam energy (Au+Au, $E_{lab}$= 0.1 A GeV - 40 A GeV, $b$=6-9 fm). While $v_4$ is always positive, $v_2$ developes a negative sign due to the onset of squeeze-out. Nevertheless, even in this geometry $v_4=v_2^2$ for both integrated equations of state.

\begin{figure}[t]	%       -----------------------------------------
\includegraphics[width=0.5\textwidth]{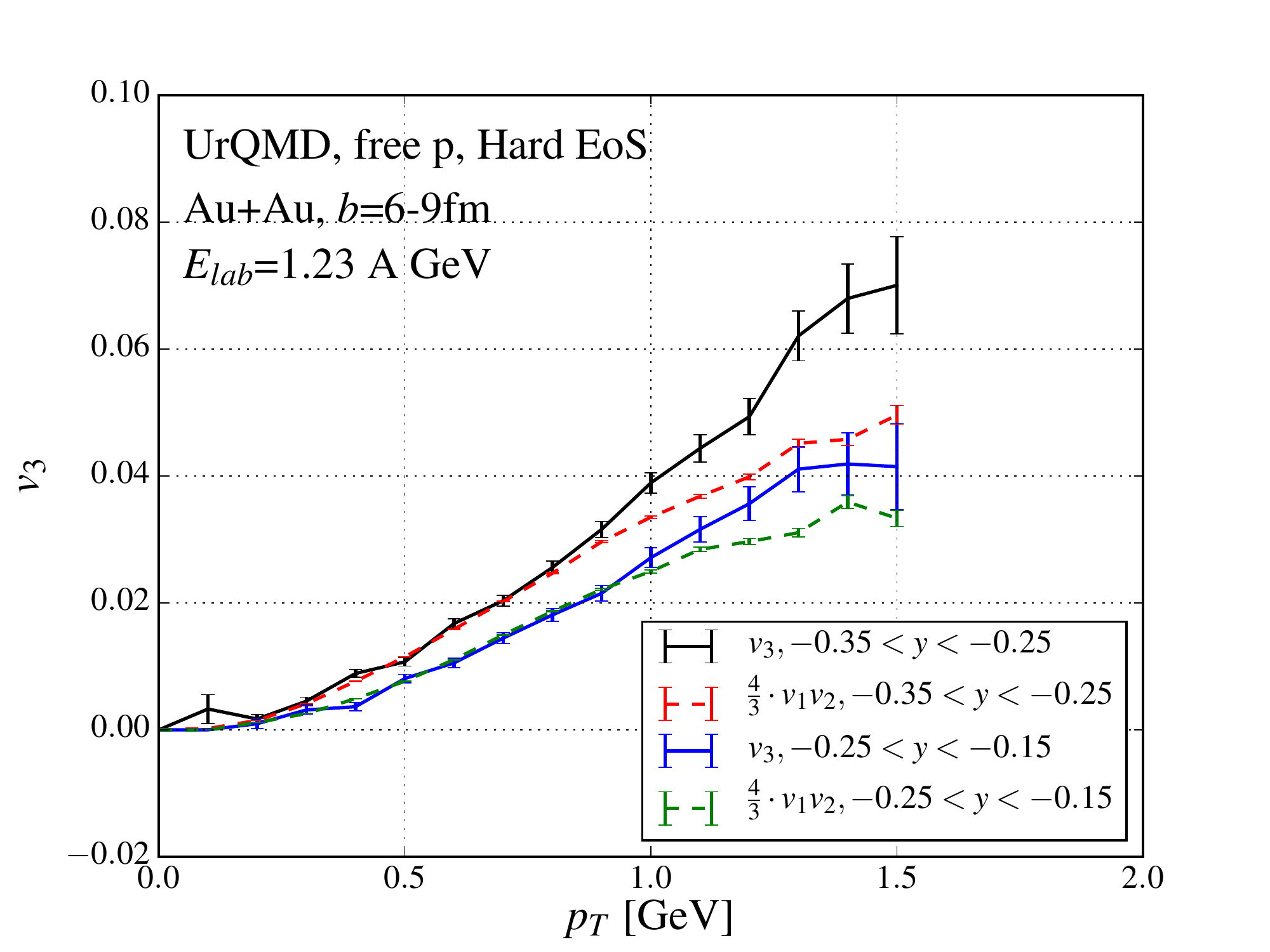}
\caption{[Color online] Flow of free  protons in Au+Au collisions ($b=6-9$ fm) as a function of transverse momentum for different rapidity windows  at a fixed-target beam energy of 1.23 AGeV. The lines indicate the UrQMD calculations for a hard equation of state. }\label{f22}
\end{figure}

\begin{figure}[t]	%       -----------------------------------------
\includegraphics[width=0.5\textwidth]{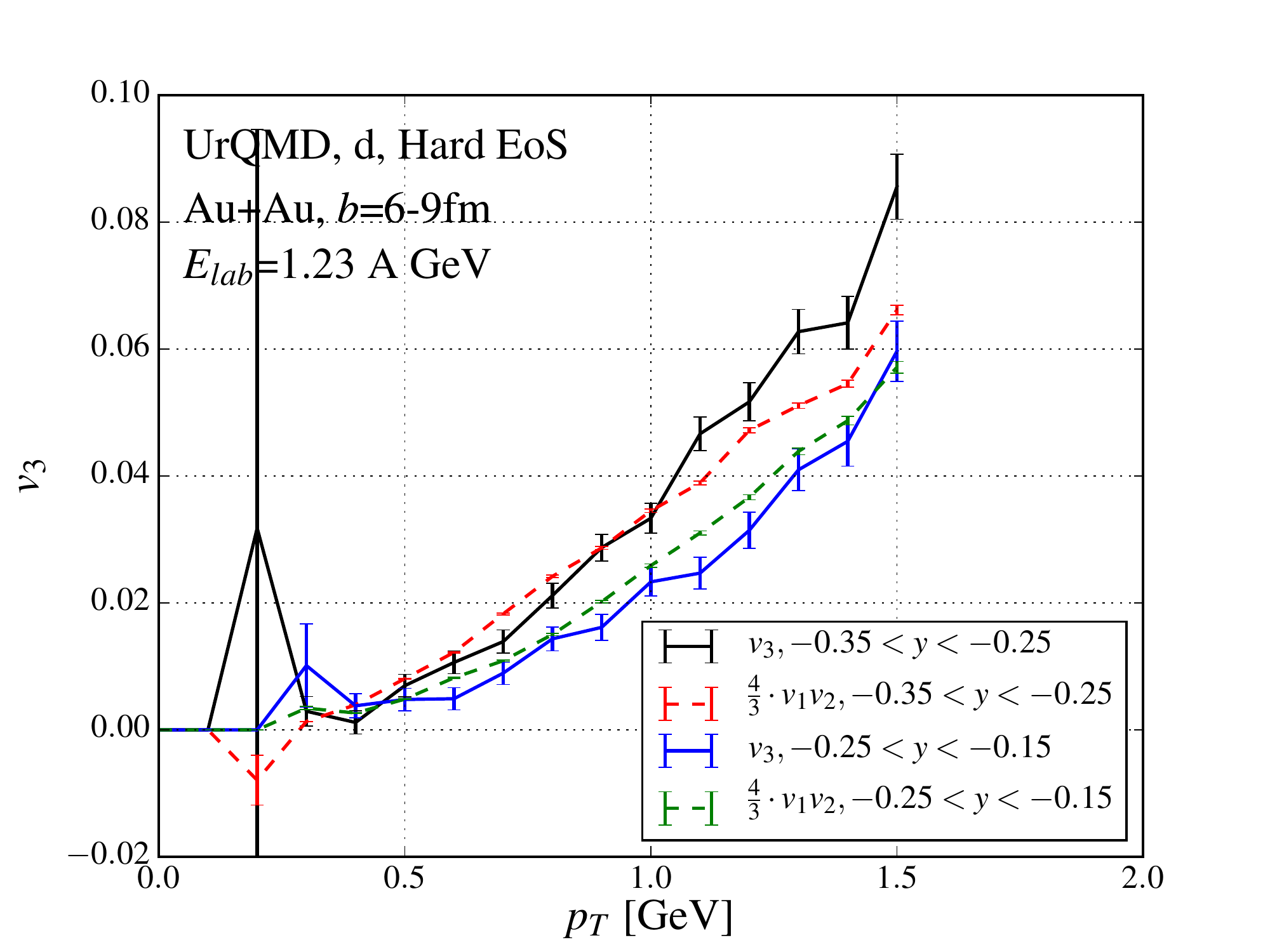}
\caption{[Color online] Flow of deuterons in Au+Au collisions ($b=6-9$ fm) as a function of transverse momentum for different rapidity windows  at a fixed-target beam energy of 1.23 AGeV. The lines indicate the UrQMD calculations for a hard equation of state. }\label{f23}
\end{figure}

Finally, we investigate flow correlations between $v_1$, $v_2$, and $v_3$. In Figs. \ref{f22} and \ref{f23} we show the protons and deuterons triangular flow $v_3$ in comparison to $\frac{4}{3}v_1v_2$ as function of transverse momentum.

One clearly observes the triangular flow of both protons and deuterons is intimately connected to the directed and elliptic flow. This supports the suggestion that all three flow components emerge from the same time dependent underlying geometry, but without correlations to initial state fluctuations.

\section{Summary}
We presented a first transport model study of the first four flow harmonics of deuterons for Au+Au collisions at a beam energy of 1.23 A GeV and a centrality of 20\%-30\%. In addition we have updated our predictions for the proton flow harmonics. The UrQMD model, with a hard momentum independent EoS, gives a very good description to the preliminary experimental data, as measured by the HADES collaboration. We have further analyzed the scaling of deuteron and proton flow and found clear indications of constituent scaling, indicating that deuterons are formed by coalescence. In addition we have explored flow correlations ($v_4\sim v_2^2$ and $v_3\sim v_1v_2$) and found clear scaling behavior. Even more surprising the numerical value $\frac{v_4}{v_2^2}=\frac{1}{2}$ suggests an ideal fluid expansion of the system.

\section{Acknowledgments}
The authors thank Behruz Kardan, Manuel Lorenz and Christoph Blume for helpful and inspiring discussions.
The computational resources were provided by the LOEWE Frankfurt Center for Scientific Computing (LOEWE-CSC) and the FUCHS-CSC. This work was supported by  HIC for FAIR and in the framework of COST Action CA15213 THOR and the DAAD-TRF collaboration.  

%%%%%%%%%%%%%%%%%%%%%%%%%%%%%%%%%%%%%%%%%%%%%%%%%%%%%%%%%%%%%%%%%%%%%%%%%%%%%%% 
\end{document}